\documentclass{aastex701}
\usepackage{amsmath}
\usepackage{booktabs}
\usepackage{siunitx}
\usepackage[font=small]{caption}
\usepackage{CJKutf8}

\usepackage{bm}

\setcitestyle{notesep={,},round,aysep={},yysep={,}}

\begin{document}
\title{Detection and classification of astronomical sources with Astro-RetinaNet in crowded stellar fields}
\correspondingauthor{Chao Liu}
\email{liuchao@bao.ac.cn}

\author[0009-0001-0987-9942]{Yibo Yan\
\begin{CJK}{UTF8}{gbsn}(闫一波)\end{CJK}}
\affiliation{National Astronomica Observatories, Chinese Academy of Sciences, Beijing 100101, P.R. China}
\affiliation{School of Astronomy and Space Science, University of Chinese Academy of Sciences, Beijing 100049, China}
\email{ybyan@bao.ac.cn}

\author[0000-0002-1802-6917]{Chao Liu\
\begin{CJK}{UTF8}{gbsn}(刘超)\end{CJK}}
\affiliation{National Astronomica Observatories, Chinese Academy of Sciences, Beijing 100101, P.R. China}
\affiliation{School of Astronomy and Space Science, University of Chinese Academy of Sciences, Beijing 100049, China}
\affiliation{Institute for Frontiers in Astronomy and Astrophysics of Beijing Normal University, Beijing, 100875, China}
\affiliation{Zhejiang Lab, Hangzhou, 311121, China}
\email{liuchao@bao.ac.cn}

\author[0000-0002-3651-5482]{JiaDong Li\
\begin{CJK}{UTF8}{gbsn}(李佳东)\end{CJK}}
\affiliation{Max-Planck-Institut für Astronomie, Königstuhl 17, D-69117 Heidelberg, Germany}
\email{jdli@mpia.de}

\author[0000-0002-9847-7805]{Feng Wang\
\begin{CJK}{UTF8}{gbsn}(王锋)\end{CJK}}
\affil{Center For Astrophysics, Guangzhou University, Guangzhou, Guangdong 510006, China}
\affil{Great Bay Center, National Astronomical Data Center, Guangzhou, Guangdong 510006, China}
\email{fengwnag@gzhu.edu.cn}
\begin{abstract}
Upcoming next-generation sky surveys will detect large number of faint objects with magnitudes larger than 25. 
When objects are crowded within a limited a field of view, blending becomes unavoidable.
Blending leads to the omission of many sources during photometry in these fields, which cause an underestimates of tens of percent in crowded fields, and remains a major challenge for existing source-extraction techniques.
Although artificial neural networks had shown promising results in the detection and classification in wide-field surveys, they often fail with severely blended stars.
We developed a robust deep learning model, Astro-RetinaNet, based on the Retinanet algorithm to detect and classify blended sources in single-band astronomical images.
After training and evaluating the performance of our network on simulated images, we find precision of 0.96, 0.89,0.70, 0.50,0.75 for single star, 2-star, 3-star, 4-star and 5-or-more star blending cases, respectively, with star number density $\sim$22000 stars per $\rm arcmin^2$.
We compare our method's detection capability and completeness both on CSST simulated NGC 2298 images and HST observed M31 images. 
In crowded and non-crowded stellar fields of simulated NGC 2298, our results show that the model can recover $82\%$ and $95\%$ sources respectively at magnitude ($i$ band) of 25, while for SExtractor and Photutils the completeness reduces to $20\%, 59\%$ and $60\%, 88\%$ respectively. 
In the M31 case, as faint as 27 magnitude ($F814W$) in a crowded field, Astro-RetinaNet detects 2,224 sources, significantly outperforming Photutils and SExtractor by factors of 3.4 and 7.1, respectively.
\end{abstract}


\section{ Introduction }\label{sec:intro}
In modern astronomical research, large-scale survey observations are indispensable, generating massive datasets that significantly advance our understanding of the universe's structure, evolution, and fundamental phenomena like dark energy. 
Projects such as the ongoing Dark Energy Survey \citep[DES,][]{DES} and Hyper Suprime-Cam Subaru Strategic Program \citep[HSC,][]{HSCSSP}, alongside upcoming initiatives like the Legacy Survey of Space and Time \citep[LSST,][]{Ivezic_2019}, the Euclid mission \citep{Euclid}, the Wide Field Infrared Survey Telescope \citep[WFIST,][]{WFIST}, and the Chinese Space Station Survey Telescope \citep[CSST,][]{CSST1,csstcollaboration2025,Gong_2019}, deliver high-resolution, wide-ranging, and deep observations that drive breakthroughs in cosmology and astrophysics.

Despite advancements in resolution and depth from these surveys, observing crowded stellar fields, such as the centers of globular clusters (GCs) and galactic bulge, remains challenging due to blending effects. 
For instance, in the globular cluster (GC) NGC 4472 at 17.14 Mpc, the stellar number density at 4–5 effective radii ($R_{\rm eff}$) reaches up to $10^4$ stars per square arcsecond \citep{Anand_ngc4472}, exceeding the resolution limits of current ground-based and space-based telescopes (typically better than 0.5 arcsec).
Even in the relatively sparse field of $\omega$ Centauri, with a stellar density of approximately $10^2$ stars per square arcsecond \citep{omega_p1}, only 30 stars within a 90 arcsecond radius have proper motion measurements with sufficient precision \citep{omega_p2}. Blending occurs when the Point Spread Functions (PSFs) of multiple sources overlap significantly on the detector, rendering them indistinguishable and hindering the study of the central properties of GCs and nearby galaxies.

To mitigate blending, deblending techniques aim to reconstruct the properties of individual sources from blended observations. 
These techniques are broadly categorized into traditional algorithms, such as PSF fitting and aperture photometry, and modern machine learning (ML) based approaches. Traditional methods, such as Source-Extractor \citep[SExtractor,][]{Bertin_1996}, use isophotal analysis to assign flux within a pixel to a single source, while the Multiband Probabilistic Cataloging method \citep{Feder_2020} employs a hierarchical Bayesian model for multiband source detection and deblending. 
Similarly, Multiband morpho-Spectral Component Analysis Deblending Tool \citep[MuSCADeT,][]{MuSCADeT} leverages morphological component analysis to separate blended gravitational lenses, and Source separation in multiband images by Constrained Matrix Factorization \citep[SCARLET,][]{scarlet} uses matrix factorization to eliminate crowded extragalactic scenes. 
ML-based approaches, such as Morpheus \citep{Hausen_2020} and Blend2mask2flux \citep{blend2mask2flux}, utilize U-Net architectures, a U-shaped convolutional neural network design \citep{Ronneberger_Unet} for pixel-level segmentation and simultaneous detection-deblending tasks.
Astro R-CNN \citep{astro-rcnn} employs Mask R-CNN \citep{mask-rcnn}, an extension of the Region-based Convolutional Neural Network (R-CNN) framework, for multi-band source detection and classification.
Additionally, modified Super Resolution Generative Adversarial Networks  (SRGAN) methods \citep{modified_scalet} address source deblending by leveraging generative models to reconstruct pixels obscured by overlapping sources.

However, both traditional and ML-based deblending methods face significant limitations, particularly in crowded stellar fields. Traditional algorithms, such as SExtractor with PSF Extractor \citep[PSFex,][]{psfex}, rely on single- or multi-band flux modeling but are highly sensitive to configuration parameters, which are challenging to optimize \citep{Hedges_2021}. Moreover, these methods depend on prior source detection to determine the number of sources within a blended region, a task that becomes increasingly difficult with higher stellar densities \citep{Vaisanen_2001, Runjing_Liu_2022}. ML-based deblenders, while promising, are often trained on simulated images tailored to specific surveys, leading to potential generalization issues when applied to real data with varying observational conditions, such as PSF shape, image resolution, or sky brightness \citep{Dawson_2015}. These challenges highlight the absence of a universal deblending solution, underscoring the need for robust methods to fully exploit the potential of current and future survey data in crowded fields.

In this study, we develop a deep learning technique based on RetinaNet \citep{Tsung-YiLin_2017} to detect and classify the blended stars.
Rather than addressing the complex case of blending galaxies, we focus on resolving blended stellar sources. 
Our model directly categorizes detections into single stars and blended units, and then further gives an estimation of the number of stars within a blending unit. 
The rest of the paper is organized as follows.   
The description of simulation pipeline and data normalization method are outlined in section \ref{sec:method}. 
In section \ref{sec:data training}, we introduce the RetinaNet framework and present our training result and model performance. In section \ref{sec:application}, we apply the model both on CSST simulations and the Hubble Space Telescope (HST) observational data and compare the model's detection capability and completeness with SExtractor and Photutils. 
Then we discuss the shortage and limitation  of the model in section \ref{discussion}. 
We summarize the main conclusions and outline future directions in section \ref{sec:conclusion}.

\section{Data Preparation}\label{sec:method}
We utilized the GalSim \citep{Rowe2015121} software package to simulate stellar fields with diverse densities and various blending scenarios. 
GalSim is an open-source toolkit for simulating astronomical objects, including stars and galaxies, using a variety of parametric models. 
Based on a specifically designed mock star catalog and by integrating GalSim with the instrumental PSF models of CSST, we generate simulated image data of crowded fields as they would be observed by the CSST.

In addition to the tools for simulating images, we also need to construct a mock catalog with sufficient density to simulate blending scenarios in crowded stellar fields. The catalog should include not only the coordinates and magnitudes of each source but also their corresponding blending categories. Therefore, the preparation of simulation data can be divided into two parts: mock catalog generation and image simulation.

\subsection{Mock Catalog}
For subsequent modeling, we constructed a mock stellar catalog containing the coordinates, photometry, and error.
Several key factors are considered in the context of blending scenarios. First, the flux difference between stars within each blending unit should not be too large because this would cause fainter stars to be obscured by brighter stars, making it undetectable. 
Second, the pixel distance between stars within a blending unit should be small enough to keep them inseparable. 
In addition, overlap between different blending units should be prevented. 

Under these constraints, simulating the coordinates and photometry of each star requires careful consideration of source interactions to ensure realistic and physically meaningful configurations. 
In particular, when two or more stars fall within the same pixel and exhibit extreme flux contrasts, the addition of the fainter component does not introduce a measurable modification to the PSF profile on the detector,  as its contribution remains below the noise level of a single resolution element. 
Although such configurations are formally blended in a geometric sense, the resulting PSF perturbations are negligible under typical undersampling conditions and are therefore treated as single-star. 
Accordingly, we adopt a definition of blending: only stars whose inclusion produces a detectable change in the PSF morphology on the CCD are treated as blended components in the simulation.

\begin{figure}[hbpt!]
\centering 
\includegraphics[width=0.45\linewidth]{ 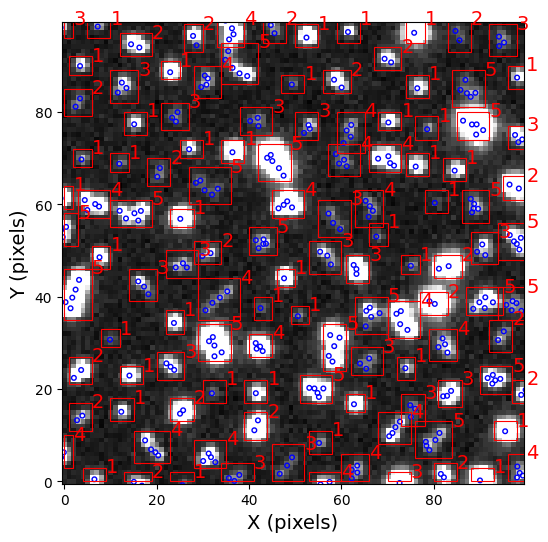} \hspace{-0.2cm}
\captionsetup{font=small} 
\caption{Typical example of GalSim CSST training set. Pixel coordinates and magnitudes were sampled from Eq. \eqref{eq1} and Eq. \eqref{eq:2}. 
The pixel coordinates for each star are denoted by blue circles, and their corresponding bounding boxes are represented by red squares, both directly labeled. The numerical value appearing in the top-right corner of each box indicates the blending type (labels).
}\label{fig:Blending_stamp} % \ref{fig:HID}.
\end{figure}

The simulation parameters were constrained by the instrumental characteristics and astrophysical requirements of CSST. 
At first, we begin with a simple case, limiting the stellar number of a blending unit to a maximum of five, as a higher count complicates classification. 
Empirical findings show that setting the center-to-center separation between units to 8 pixels . 
Given that a typical blending unit extends over a scale of $\sim5$ pixels, this spacing corresponds to an overlap of about $4/25$ of the unit area, providing a balance between realistic crowding and reliable separability.
Considering the CSST's pixel scale of 0.074 arcsec, this configuration corresponds to a stellar number density of $\rho_N \in [3,20] \ \rm{stars \ arcsec^{-2}}$.
As we can see in Figure. \ref{fig:Blending_stamp}, each blending unit can be treated relatively isolated within the bounding box, ensuring minimal overlap between adjacent units. 

Regarding the magnitude range, the CSST detectors are operated within 18 to 25 mag. 
To evaluate the performance of the model on stars fainter than the nominal detection limit, we restricted the simulated stars to the range of 20 to 26 mag.
Finally, to prevent the photometric obscuration of faint stars by brighter neighbors, the maximum flux difference within any blending unit was limited to 2.5 mag, corresponding to a factor of 10 in flux. 
Exceeding this limit would cause the faint star to be undetectable, leaving the blending system unresolved.

Here, we outline our approaches to address these challenges.
\begin{itemize}
\item[1).] The first step is to randomly scatter a set of points, which we refer to as "base stars." 
They serve as the initial members of a blending unit because they form the foundational seeds for the subsequent blending growth process. 
At the same time, the magnitudes of these base stars are drawn from a luminosity function Eq. \eqref{eq1}. 
\begin{gather}\label{eq1}
    N(m) = a\cdot10^{b\cdot m} .
\end{gather}
where $N(m)$ is the number of base stars at magnitude $m$, and $a,b$ are the parameter index.
Here we set the minimum distance between base stars at least 8-pixel, which ensures that the blending units do not overlap each other (as shown in the top-left of Figure. \ref{fig:pos_df}. 
Notably, the magnitudes of these seeds are the brightest in each blending unit, whereas the brightnesses of the other member stars decrease progressively.

\item[2).] After these base stars are properly placed, the next step is to construct 2-star blending category by randomly selecting a subset of base stars and placing a fainter companion around each of them.
Here,  we use $x_c,y_c, f_c, r_c, m_c$ to represent the pixel coordinates, flux (in electrons), flux-radius (in pixels) and magnitude of the central star, respectively.
The photometric zero point ($zpmag$), exposure time ($exptime$, in seconds), and CCD gain ($gain$, in $\mathrm{e^-\,ADU^{-1}}$) are adopted to convert magnitudes into detector counts.
This star is the one around which a neighboring blending member will be placed, and now the base star is treated as a central star.
The position and flux of the second star are then assigned using Eq. \eqref{eq:2} below:
\begin{equation}\label{eq:2}
\begin{aligned}
    m_2 &\sim {\rm {U}} (m_c, m_c + 2.5) ,\\
    f_2 &= 10^{-0.4 \cdot (m_2 - zpmag)} \cdot  exptime \cdot  gain ,\\
    dist_{max} &= r_c(f_c = f_2\cdot  0.5) + r_2(f_2 = f_2\cdot  0.5) ~(\rm pixels),\\
    dist_{min} &= 0.5 ~(\rm pixels).
\end{aligned}
\end{equation}
where $m_2$ denotes the magnitude of the second star, uniformly sampled from the interval $(m_c, m_c+2.5)$. The variables $f_2$ and $r_2$ represent the flux and flux radius of the second star, respectively.
We define the maximum separation, $dist_{max}$ in Eq. \eqref{eq:2}, as the distance at which the peak flux of the member star falls to $50\%$ of the flux profile of the central star. 
If the separation of two stars exceeds $dist_{max}$, it means that we can split them clearly, and they will be treated as single stars, respectively. 
In contrast, the minimum separation is set to 0.5 pixels. Below this threshold, the combined flux profile of the two stars is
indistinguishable from that of a single star, and they together will be treated as a single star. 
Finally, the 2-star blending category is represented by the green point in the top-right panel of Figure. \ref{fig:pos_df}. 
And the remaining base stars (black points) that are not used to construct 2-star blending cases are classified as single-star samples.

\item[3).] Next, we start by inserting the tertiary star into the randomly chosen 2-star blending unit.
Unlike in the previous step, the base star no longer serves as the central star; instead, the secondary star now acts as the central star. Like the operation on the Step 2, the magnitude of the third star $m_3$ will be sampled in the interval $(m_2, m_c+2.5)$.
Here, the lower limit of the magnitude range is set by the magnitude of the second star $m_2$, while the upper limit remains tied to the base-star magnitude $m_c$.
Meanwhile, the position of the tertiary star will be uniformly sampled within the radius range $(0.5, dist_{max})$, centered on the secondary star.

This growth strategy prevents all member stars from clustering around the base star, thereby promoting a variety of blending morphologies (e.g., bar-like, circular, or L-shaped configurations). 
It also ensures that the flux ratio within the blending unit does not exceed a factor of 10.
We use a red point to represent the scenario after the third member star joined the blending unit, with the result shown in the bottom-left panel of Figure. \ref{fig:pos_df}.

\begin{figure}
\centering 
%\hspace{-0.4cm}
\includegraphics[width=0.6\linewidth]{ 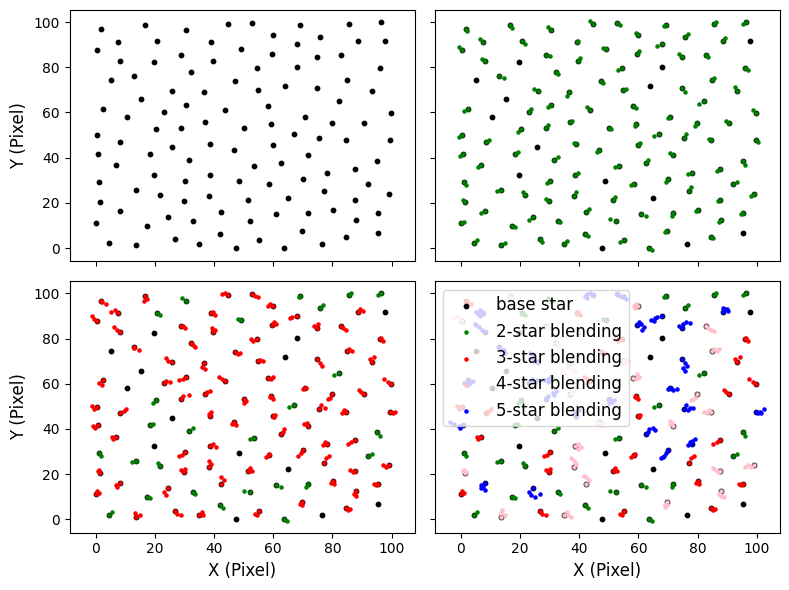}   \hspace{-0.5cm}
\captionsetup{font=small}
\caption{Mock stellar catalog generation process (Top-left $\rightarrow$Top-right $\rightarrow$ Bottom-left $\rightarrow$ Bottom-right) in each image. Top-left: Initial base-star coordinates. These scattered points are spaced more than 8 pixels apart, with magnitudes ranging from 20 to 26.
Top-right: Select some base-stars as centers. Using Eq. \eqref{eq:2} as criteria, randomly distribute 2-star blending (green dots) within the defined radius.
Bottom-left: Similarly, select some secondary stars from 2-star blending as centers and distribute 3-star blending positions (red dots) within the radius.
Bottom-right: The same principle is applied until 5-or-more-star aliased member stars are generated. Here, 4-star and 5-or-more-star are represented by pink and purple dots, respectively. \label{fig:pos_df}}
\end{figure}

\item[4).]  Similar to the 3-star blending case, the magnitude and positions of the 4-star blending and 5-or-more-star blending categories are sequentially generated in the same pattern. 
To avoid potential overlaps between these blending units due to a too large number of stars in a blending unit, we limit the number of stars per blending unit to no more than 5. 
In other words, the blending unit in the training data can be categorized into five classes. The final result is shown in the bottom right panel of Figure. \ref{fig:pos_df}.
\end{itemize}

\begin{figure}[htbp]
\centering 
\includegraphics[width=0.85\linewidth]{ 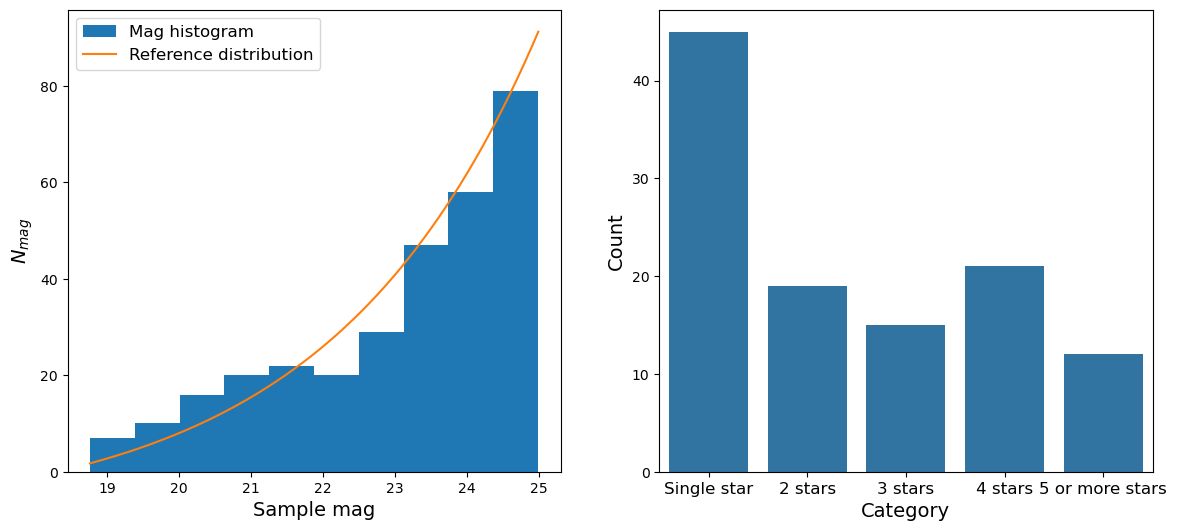}
\captionsetup{font=small} 
\caption{Left: Magnitude distribution of mock catalog, with the blue histogram showing observed power-law counts and the yellow curve indicating the  power-law probability density function (PDF). Right: A typical example of the distribution of blending categories in a single training image, where single stars (twice as frequent as blended categories) dominate to prevent misclassification, while other blending categories follow a uniform distribution.}\label{fig:mag_type_DF} % \ref{fig:HID}.
\end{figure}

In crowded stellar fields, the relative fractions of single and blended sources are not characterized by a single universal value.
In general, single stars tend to dominate in the outer regions of dense environments, whereas the blending fraction can reach up to $\sim 50\%$ in crowded regions such as the Galactic disk \citep{Kiss_l}.
In contrast, at high Galactic latitudes, single stars typically account for $80\%$–$90\%$ \citep{Kiss_l} of the detected sources and remain the dominant population.
Overall, single stars constitute the majority of sources across most observational conditions.
For simplicity, we therefore adopt an initial single-to-blended source ratio of $2{:}1$ for each simulated image (right panel of Figure. \ref{fig:mag_type_DF}), representing a compromise between sparse and crowded regimes.
Empirically, we find that increasing this ratio further (e.g., to $3{:}1$) significantly degrades performance in crowded regions, with most blended sources incorrectly identified as single stars.
In addition, we parameterized the magnitude-density relation using power-law derived from the PHAT photometric data of M31 (left panel of Figure. \ref{fig:mag_type_DF}).

\subsection{Image generation}

Following the construction of the mock catalog,
we generate a simulated crowded field image by convolving it with the CSST $g$-band PSF \citep{weichengliang2025}. 
The image simulation process, analogous to catalog generation, is divided into two key steps:
\textbf{(i)} generating a noise-free image through convolution, and
\textbf{(ii)} adding various noise components.
A detailed description of each step is provided below.

The image simulation begins with a convolution operation for each star on a $100 \times 100$ pixel stamp, corresponding to an area of approximately 7.4 arcsec $\times$ 7.4 arcsec. 
Based on the CSST's observational parameters (Table \ref{tab:csst_params}), we first convert the magnitudes from the mock catalog into flux values.
Since stars are treated as point sources, their intrinsic flux distribution can be represented by a Delta function.
Convolving this Delta function with the instrumental PSF profile, we obtain clean stellar profiles. 
Then the stellar profiles of member stars in a blending unit are generated and combined onto an individual stamp. 
Each blending unit is processed separately. 
Finally, by co-adding all these individual stamps, a noise-free simulated image of the blended stellar fields is generated.

\begin{table}[htbp]
  \centering
  \caption{CSST observational parameters}
  \label{tab:csst_params}
  \footnotesize % 缩小字体
  \setlength{\tabcolsep}{3pt} % 减小列间距（默认6pt）
  \renewcommand{\arraystretch}{0.85} % 减小行高
  \begin{tabular}{@{}llll@{}}
    \toprule
    \textbf{Parameter} & \textbf{Description} & \textbf{Value} & \textbf{Unit} \\
    \midrule
    \textbf{Exposure time} & Single image duration & 150 & s \\
    \midrule 
    \textbf{Filter} & Band & g, r, i & -- \\
    & Central wavelength & 475, 625, 775 & nm \\
    & Bandwidth & \num{150}, \num{140}, \num{140} & nm \\
    \midrule
    \textbf{Noise} & & & \\
    \quad Sky level & Poisson noise & $2\times10^3$ & e/s/pixel \\
    \quad Readout noise & Gaussian noise & $\mu=0$, $\sigma=\num{5.2}$ & e/s/pixe; \\
    \quad Dark current & Thermal noise & \num{0.8} & e/s/pixel \\
    \midrule
    \textbf{CCD} & Pixel scale & \num{0.074} & arcsec/pixel \\
    & Saturation level & \num{65535} & e$^-$/s \\
    & Gain & \num{1.0} & e$^-$/ADU \\
    \bottomrule
  \end{tabular}
\end{table}

In terms of noise sources, we consider the noise distributions derived from CSST observational parameters in Table \ref{tab:csst_params}. 
These include Poisson noise from both the sky background and stellar fluxes, Gaussian readout noise from the CCD, and dark current and its associated shot noise.
Adding these components to the noise-free stamp yields a realistic average noise level. 

Figure. \ref{fig:Blending_stamp} shows a simulated blending image of a crowded stellar fields with the above method.
The entire dataset comprises 5,000 images, containing approximately $10^6$ stellar targets in total. 
The pixel area of these blending units varies depending on the blending category. 

Additionally, since the base-star separation is set to 8 pixels, significant overlap between blending units is prevented. 
Even when overlap occurs, it only affects the border regions and does not affect the ground truth of the blending types. 
This is illustrated in Figure. \ref{fig:Blending_stamp}, we can find some blending units exhibit border intersection, but no sources are distributed within the overlapping region, ensuring the precision of labels in the dataset.

\subsection{Data Normalization and Cleaning}\label{subsec: data Normalization}
In our blending tasks, the primary objective is not only the source detection but also the accurate classification of blended structures. 
To address this, our approach is designed to conserve the flux profiles of all sources within a blending unit and enhance the signal-to-noise ratio without affecting their relative brightness differences.

To do so, We first estimate the sky background distribution using SExtractor \citep{Bertin_1996}, take its global mean $F_{bkg}$, and replace the pixel values below the threshold with $F_{bkg}$ to suppress background noise in low SNR regions. The corrected flux of pixel $(i,j)$ in the background-subtracted image can be written as:

\begin{gather}\label{eq4}
    F_{ij} =max(F_{ij}^{raw}, F_{bkg}) .
\end{gather}

where $F_{ij}^{raw}$ is the original flux value at raw $i$ and column $j$.  Based on the physical limits of the CCD detector (maximum ADU value $2^{10}-1=65535$) and observational parameters (gain $G$, exposure time $T$), the theoretical maximum pixel value is calculated as $F_{max} =65535\times G/T ~~~[e^-/s]$.

Then, pixel values can be linearly scaled to the interval [0,1]. Here we define the normalized flux as 
\begin{gather}\label{eq6}
    F_{ij}^{norm} = \frac{F_{ij}- F_{bkg}}{F_{max}-F_{bkg}}.
\end{gather}
where $F_{ij}^{norm}$ represents the  normalized flux in the pixel $(i,j)$. This approach preserves the dynamic range of astrophysical signals, while eliminating background offsets and standardizing the data to a unified scale.

\section{Astro-RetinaNet and  Its Performance}\label{sec:data training}
This section details the model framework and training procedure, the training parameters used, and the analysis of the results through confusion matrices and Precision-Recall (PR) curves.

\subsection{The Astro-RetinaNet}\label{sec:astro-retinanet}

To address the stellar blending problem in a crowded field, our method is built on RetinaNet \citep{Tsung-YiLin_2017}, named for its dense sampling of object locations in an input image and well-suited for this task. 
RetinaNet is a unified network architecture comprising a backbone network and two subnets specific to tasks.  

\begin{figure}
\centering 
%\hspace{-0.4cm}
\includegraphics[width=0.9\linewidth]{ 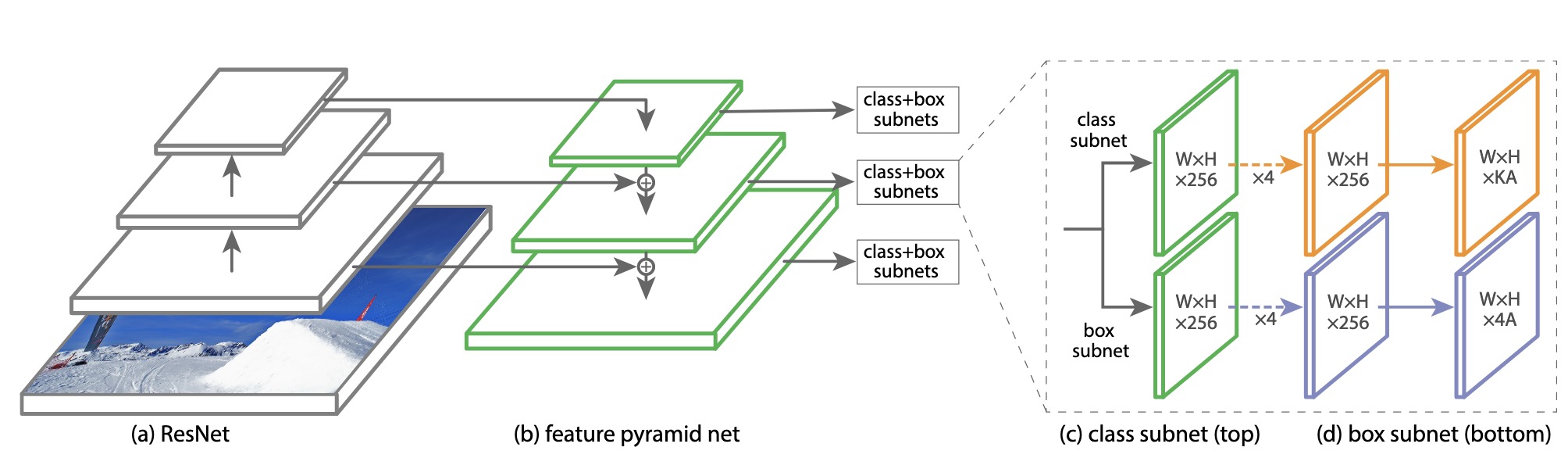}   \hspace{-0.5cm}
\captionsetup{font=small} 
\caption{Reprinted from \citep{Tsung-YiLin_2017}, with permission from Tsung-Yi Lin.  The one-stage RetinaNet network architecture uses a Feature Pyramid Network (FPN) backbone on top of a feedforward ResNet architecture (a) to generate a rich, multi-scale convolutional feature pyramid (b). To this backbone RetinaNet attaches two subnetworks, one for classifying anchor boxes (c) and one for regressing from anchor boxes to ground truth object boxes (d). The network design is intentionally simple, which enables this work to focus on a novel focal loss function that eliminates the accuracy gap between our one-stage detector and state-of-the-art two-stage detectors like Faster R-CNN with FPN while running at faster speeds..}\label{retinanet} 
\end{figure}

Both subnets adopt a simple design that facilitates one-stage dense object detection, as illustrated in Figure. \ref{retinanet}.
In general, object detectors can be categorized into one-stage and two-stage methods. 
Two-stage object detectors \citetext{e.g., Fast R-CNN~\citealp{F-RCNN_Ren}; Mask R-CNN~\citealp{mask-rcnn}; R-CNN~\citealp{RCNN_girshick}} first generate region proposals and then perform classification and localization, whereas one-stage object detectors \citetext{e.g., YOLO~\citealp{YOLO_Redmon}; SSD~\citealp{SSD_Liuwei};RetinaNet~\citealp{Tsung-YiLin_2017}} directly predict object classes and bounding boxes in a single step.

A key characteristic of RetinaNet is its use of focal loss, a modified cross-entropy (CE) loss that focuses training on hard negative examples. 
We introduce the focal loss starting from the CE loss which is defined as:
\begin{gather}
    \label{CE loss}
    CE = - \sum^{K}_{i} t_{i} \log(s_{i}).
\end{gather}
where  $t_i$ and $s_i$ are the ground truth and the score for each class $i$ in K classes. 
The common method for addressing class imbalances is to introduce weighting factor to balance the importance of positive/negative examples, it does not differentiate between easy/hard examples. 
Instead, they propose to reshape the loss function to down-weight easy examples and thus focus training on hard negatives.

With this strategy, they claim to solve the problem of class imbalance by making the loss implicitly focus in those problematic classes.
We define the focal loss as:
\begin{gather}
    \label{CE loss}
    FL = - \sum^{K}_{i} (1-s_i)^{\gamma}t_{i} \log(s_{i}).
\end{gather}
where $(1-s_i)^{\gamma}$ is the modulating factor to CE loss, with the focusing parameter $\gamma \geq 0$.
Following their experiments result, we adopt the same parameter setting and set $\gamma = 2$ in our experiments.

This innovation enables RetinaNet to match the inference speed of earlier one-stage object detectors while surpassing the accuracy of contemporary two-stage detectors.
Our code in this work is developed in Python language, using the Pytorch library \citep{pytorch_cite} to construct the framework.

Our Astro-RetinaNet implementation uses the depth 50 ResNets \citep[ResNet-50;][]{he2015deepresiduallearningimage}  with a Feature Pyramid Network (FPN) built on top as its backbone. ResNet-50 is a feature extractor, wherein earlier layers detect low-level features (e.g., image corners or edges) and later levels detect high-level features (e.g., blending stars) using residual learning. 
The FPN addresses the limitation of single-layer ResNet features by augmenting the backbone with a top-down pathway and lateral connections, thereby constructing a rich, multi-scale feature pyramid from a single-input image \citep{lin2017featurepyramidnetworksobject}, as illustrated in Figure. \ref{retinanet}(a)-(b).

The object classification subnet and the box regression subnet share a common structure, a small fully convolutional networks (FCN) attached to each FPN levels, but use separate parameters. 
The classification subnet predicts the probability of the existence of an object at each spatial position for each of the anchors and K object classes, where K is the number of object classes.
When each anchor is assigned, a 4-vector output that can generate a suggested shifted bounding box to place the object at its center. 
The design of the box regression subnet aims to predict the relative offset between the anchor and the ground truth box.

\subsection{Training}
The model was trained on a dataset of 4,500 simulated images using an NVIDIA Tesla V100 GPU. 
The ResNet-50-FPN backbone was pre-trained on the ImageNet1k dataset \citep{Imagenet_deng}, and optimized using the Adam optimizer \citep{kingma2017adam}. 
The update rule of parameter $\theta$ is defined as:
\begin{gather}
    \label{Adam_optimize}
    \theta_{t} = \theta_{t-1} - \frac{\eta \cdot \hat{m}_t}{\sqrt{\hat{v}_t}+\epsilon}.
\end{gather}

Here, we set the initial learning rate $\eta=1\times10^{-2}$, with momentum coefficients $\beta_1=0.9$ and $\beta_2=0.999$ and a numerical stability constant  $\epsilon =10^{-8}$. 

\begin{figure}[htbp]
\centering 
%\hspace{-0.4cm}
\includegraphics[width=0.6\linewidth]{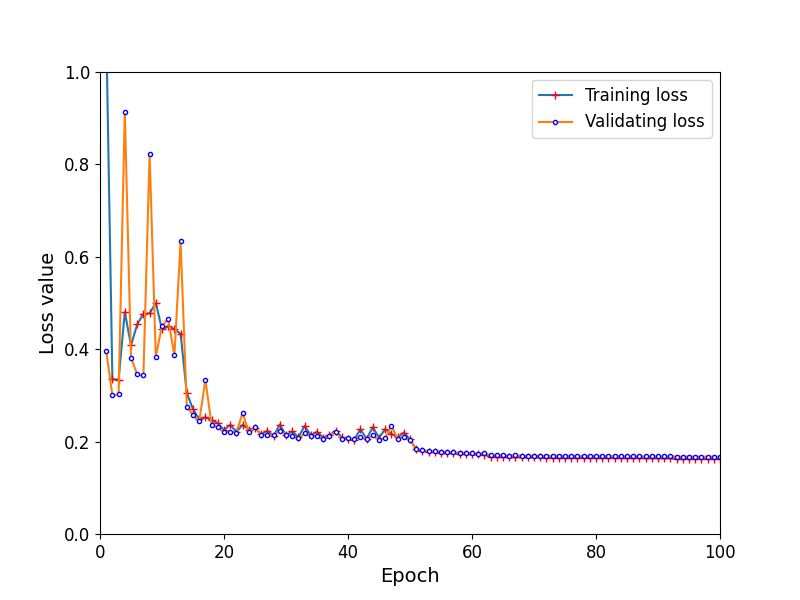}   
\captionsetup{font=small} 
\caption{Loss function curves during training (solid blue line with red circles) and validation (solid origin line with blue circles). The model converges at 50 epochs with a total training duration of 100 epochs. }\label{fig:loss_trend} 
\end{figure}

The training process ran for 100 epochs, taking approximately 4 hours, with the final composite loss converging to $L = L_{cls} + L_{box} = 0.143$, where $L_{cls}$ and $L_{box}$ denote the classification and bounding box regression losses, respectively. 
The model achieved effective convergence within this period, as indicated by the stabilization of both the training and validation loss curves (Figure. \ref{fig:loss_trend}).

\subsection{Performance analysis}\label{Performance analysis}
\begin{figure}
\centering 
\includegraphics[width=0.6\linewidth]{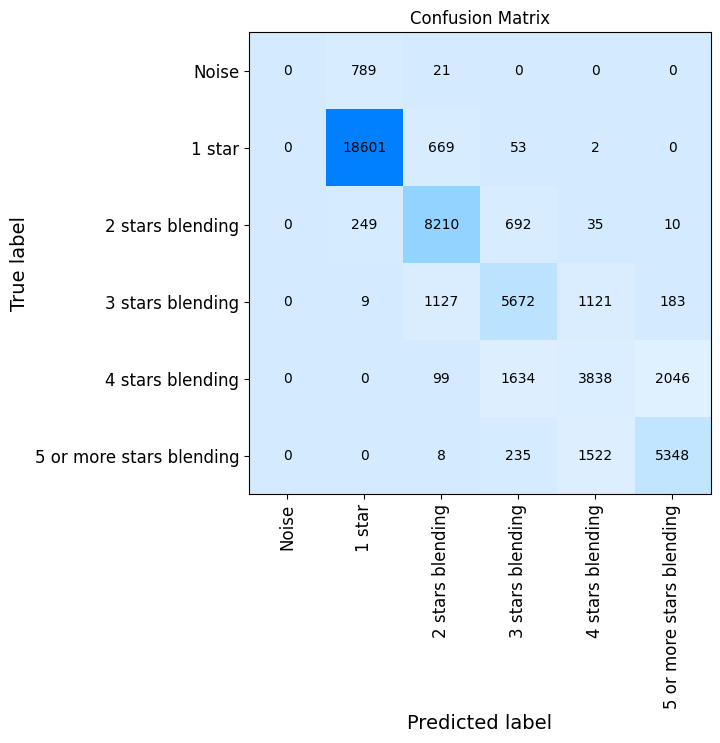}   \hspace{-0.5cm}
\captionsetup{font=small} 
\caption{Diagram of confusion matrix of classification capability tested in test data set, over 140,000 sources(at least 85\% of sources are blending stars) involved in this statistics. The darker the color in the diagonal, the better the performance of the classifier.}\label{fig:Matrix} % \ref{fig:HID}.
\end{figure}

We evaluated the model's performance on a test set that was strictly excluded from both training and validation. 
During inference, the detection results were post-processed using Non-Maximum Suppression (NMS), which iteratively removes lower scoring bounding boxes that have a high overlap with higher scoring ones, using an intersection-over-union (IoU) threshold of 0.2 to eliminate redundant detections in crowded regions.
Because our targets are predominantly small and frequently blended sources.
For such objects, modest positional offsets or partial overlaps can significantly reduce IoU, making the common threshold (0.5) overly restrictive and physically inappropriate in crowded fields.
This design ensures an unbiased estimation of the model's performance, including detection accuracy, classification precision, and false positive rates. 
To quantify the model's classification performance, we constructed an extended confusion matrix (Figure. \ref{fig:Matrix}) that incorporates a dedicated ``noise'' category to differentiate true astrophysical sources from background noise.

Based on the outcomes from the extended confusion matrix, the standard metrics of precision P (purity) and recall R (completeness) are defined as:
\begin{gather}
\label{PR_eq}
P = \frac{TP}{TP+FP},~~~~ R = \frac{TP}{TP+FN}.
\end{gather}

From the confusion matrix in Figure. \ref{fig:Matrix}, the model successfully classified all detected true sources as positive signal, resulting in a missing Detection Rate (MDR, 
the proportion of true blending units misclassified as noise by the model) 0\%, as shown by the cumulative value of zero in the noise column. 
The False Alarm Rate (FAR, the proportion of noise that was misclassified as blended sources) was approximately 1.5\%, corresponding to 810 noise signals being misclassified as positive signals. 
This low FAR confirms the model's high overall sensitivity, with only a small fraction of ambiguous noise signals remaining misclassified. 

A preliminary analysis shows that the model achieves a precision of $96.3\%$ (18,601/19,325) in identifying single stars; however, for mixed categories, the overall precision only has $72.0\%$ (23,068/32,038). 
A clear trend is observed in where the classification precision decreases with increasing blending complexity: the model achieves a precision of $89.3\%$ (8,210/9,196), $69.9\%$ (5,672/8,112), and $50.4\%$ (3,838/7,617) for 2-star, 3-star, and 4-star blending, respectively.
In contrast, blending involving 5-or-more-star exhibits a higher precision of $75.2\%$ (5,348/7,113).
We attribute this anomaly to the adopted label definitions.
The ``5-or-more-stars'' category was defined to consist of systems containing five or more stars, whose diverse spatial configurations allow the model to capture common features more effectively.
In contrast, the "4-stars" category was strictly limited to systems with exactly four stars, making their structures more similar to both 3-star and 5-or-more-star blending cases and thereby increasing classification ambiguity.
Notably, $93.5\%$ (9, 060/9, 694) of the misclassified cases differ from the true labels by only $\pm 1$ star (e.g., a 3-star blending identified as a 2- or 4-star system).
This suggests that the model can distinguish single stars from blended targets. 
For systems with higher blending complexity, the classification precision may be further improved by incorporating a Bayesian posterior framework.

\begin{figure}[htbp]
\centering 
%\hspace{-0.4cm}
\includegraphics[width=0.6\linewidth]{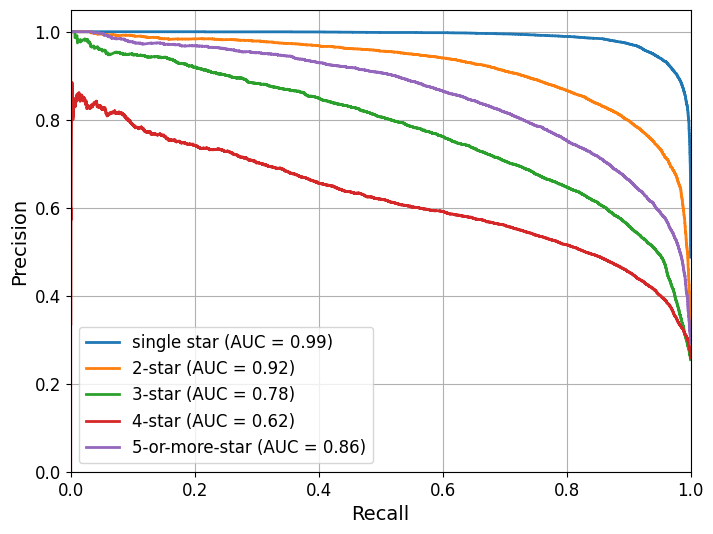}   
\captionsetup{font=small} 
\caption{ Precision-Recall (PR) curves for Astro-RetinaNet. The mean average precision (AP) score or area under curve (AUC) is calculated at an Intersection-over-Union (IoU) threshold of $\delta = 0.2$. Each solid line represents a different blending category: single stars (blue), 2-star blends (orange), 3-star blends (green), 4-star blends (red), and 5-or-more-star blends (violet).} \label{fig:PR_curve} % \ref{fig:HID}.
\end{figure}

The Precision-Recall curves in Figure. \ref{fig:PR_curve} provide further insight, particularly with respect to the area under the curve (AUC) for the blended categories. 
For blended systems except 4-star blending, the model maintains a precision above 0.6 when recall remains below 0.8. 
This indicates that while the network detects most sources , it is still hard to classify accurately the specific blending number with increased stellar counts (e.g., 4-5 stars). 

The elevated misclassification rate in these cases likely arises from two factors. 
One is the model limitation, that the current architecture may lack the capacity to disentangle overlapping flux profiles in high-density regions. 
The other is the data limitation. 
The simulated training data might oversimplify extreme blending scenarios (e.g., blending with $\geq6$ stars), leading to inadequate feature representation. 
To address these limitations, future work will focus on improving the capacity of the network and refining the simulation pipeline. 

\section{Blending Identification Using ASTRO-RETINANET}\label{sec:application}
In this section, we will mainly analyze the blending identification on CSST simulated images of NGC 2298 and real HST observations of the M31 galaxy.
In addition to the characteristics of the blending distribution, we also compare the detection capability between Astro-RetinaNet, SExtractor and Photutils. This comparative analysis evaluates the performance of the model in the following areas:
\begin{itemize}
    \item Simulated vs. real data (testing generalization capability)

    \item Sparse vs. crowded stellar fields (probing detection and classification capability)
\end{itemize}

\subsection{Blending identification and source finding in simulated NGC 2298} \label{detection comparision}

\begin{figure}[htbp]
\centering 
%\hspace{-0.4cm}
\includegraphics[width=0.85\linewidth]{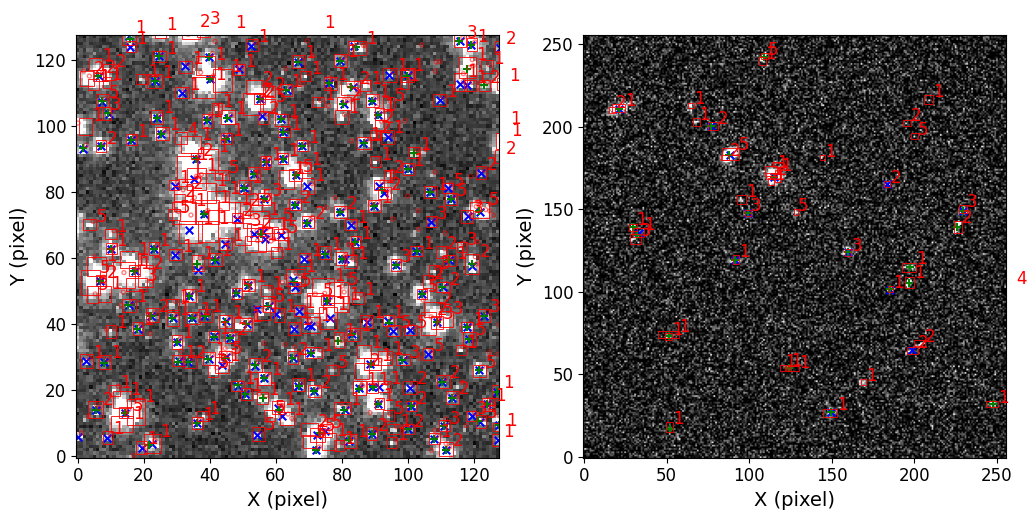}
\captionsetup{font=small} 
\caption{Detection and classification results of blended sources in central and outskirt region of the CSST-simulated NGC 2298 field using Astro-RetinaNet. Left: Astro-RetinaNet detections (red boxes), compared with SExtractor (blue `$\times$') and Photutils (green `$+$') results. Right: results of outskirt part of simulated NGC 2298 as non-crowded field, using the same symbol notation as the left panel.} \label{fig:Detection ability} 
\end{figure}

We first examine the model's performance in a crowded stellar fields, as shown in Figure. \ref{fig:Detection ability}.
The left panel displays  the detection results from Astro-RetinaNet, which are marked with red bounding boxes annotated with their predicted blending categories.
For comparison, the detection results of Sextractor and Photutils are denoted by blue `$\times$' and green `$+$' symbols, respectively. 
In crowded field, Astro-RetinaNet exhibits high sensitivity to faint candidates. 
However, this also leads to a tendency to classify some spurious signal (such as cosmic rays or background fluctuations) as positive blending signals, as can be clearly seen in the non-crowded field (right panel of Figure. \ref{fig:Detection ability}).

Comparison of classification result with ground truth reveals characteristic of error patterns:
\begin{itemize}
    \item The detected sources are frequently misclassified as single stars.
    \item The flux wings of bright stars are often misidentified as positive blended signals.
    \item The 4-star blending category is the most challenging to identify correctly.
    \item Approximately 20\% of single stars are misclassified as blended targets.
    \item Sources misclassified as the `5+' category are distributed nearly uniformly across the other true categories.
\end{itemize}

Quantitatively, the model achieves high accuracy only for single stars.
The performance degrades for 2-star and 3-star blending and is poorest for the 4-star and 5-or-more-star categories, where the classification accuracy becomes comparable to that of random classification.
One contributing factor to this behavior is class imbalance in the training set, in which the ``single-star'' class contains approximately twice as many samples as the blended categories, biasing the model toward predicting the dominant class.

To quantify the impact of the single-star fraction, we tested three class-distribution schemes across the five blending categories (1-star to 5-or-more-star):
$\rm{P}_1 = (1:2:2:2:2), \quad \rm{P}_2 = (1:1:1:1:1), \quad \rm{P}_3 = (2:1:1:1:1).$
Here, each tuple denotes the relative proportions of the single, 2-star, 3-star, 4-star, and 5-or-more-star blending classes, respectively.
We find that models trained with different single-star fractions are not highly sensitive to variations in the single-star proportion in the validation set. For example, when a model trained with the $\rm{P}_1$ distribution is evaluated on a validation set following $\rm{P}_3$, the resulting confusion matrix remains similar to that shown in Figure. \ref{fig:Matrix}, preserving a strong diagonal structure.

However, when applied to the CSST-simulated image of NGC 2298, the differences among models become more apparent.
The model trained with $\rm{P}_1$ have a precision of $75.9\%~(7,377/9,716)$ for single star class, which assigns a lower relative weight to single stars, still shows a strong bias toward classifying sources as single stars, although its single-star classification precision is lower than those of the $\rm{P}_2$ ($88.7\%~(7,736/8,714)$) and $\rm{P}_3$ ($82.3\%~(8,580/10,419)$) models. 
The $\rm{P}_3$ model yields the highest misclassification rate of blended sources into the single-star class: 71.0\% (2,603/3,659) of 2-star systems, 57.1\% (776/1,358) of 3-star systems, 42.1\% (226/537) of 4-star systems, and 24.2\% (127/524) of systems with 5-or-more stars are misclassified as single stars. In comparison, the corresponding error rates for the $\rm{P}_1$ model are 54.5\% (1,902/3,484), 39.4\% (564/1,432), 27.2\% (200/734), and 15.6\% (134/860), respectively.

Although reducing the relative weight of the single-star class indeed alleviates the tendency toward single-star predictions, it does not resolve the misclassification of blended sources as single stars. 
We believe this limitation is partly attributable to the simplicity of our simulated blended images, which do not incorporate the full range of instrumental effects present in CSST simulations. 
The resulting differences in stellar profiles may therefore contribute to classification failures.

To evaluate the model’s ability to recover the underlying stellar population, we assess the model performance by comparing the total number of stars inferred from the predictions with the ground truth counts, independent of the exact blending category.
Despite its higher misclassification rate for single stars, the $\rm{P}_3$ model exhibits superior detection performance compared to the $\rm{P}_1$ and $\rm{P}_2$ models. 
While the origin of this difference remains unclear, we adopt the $\rm{P}_3$ model for assessing stellar number recovery, since none of the tested class distributions fully eliminates the confusion between single and blended sources and $\rm{P}_3$ provides the highest detection completeness.
After excluding spurious detections, the model recovers $84.44\%$ of the true stellar population.

We next quantify how classification errors propagate into biases in stellar number estimates.
Misclassifying blended systems as single stars leads to an underestimation of the total star counts, while overestimating the blending order results in an overestimation.
After accounting for all error contributions, $25.9\% (7,615/29,404)$ of the sources lead to an underestimation of the total star counts. 
Of these, $7,437  (97.4\%)$ sources arise purely from misclassification as single stars.
In contrast, $18.6\%~(5,490/29,404)$ of the sources contribute to star-count overestimation due to overpredicted blending complexity.
Overall, the net effect is a systematic underestimation of the total stellar population, primarily driven by the bias toward single-star classification.

\begin{figure}
\centering 
%\hspace{-0.4cm}
\includegraphics[width=0.45\linewidth]{ 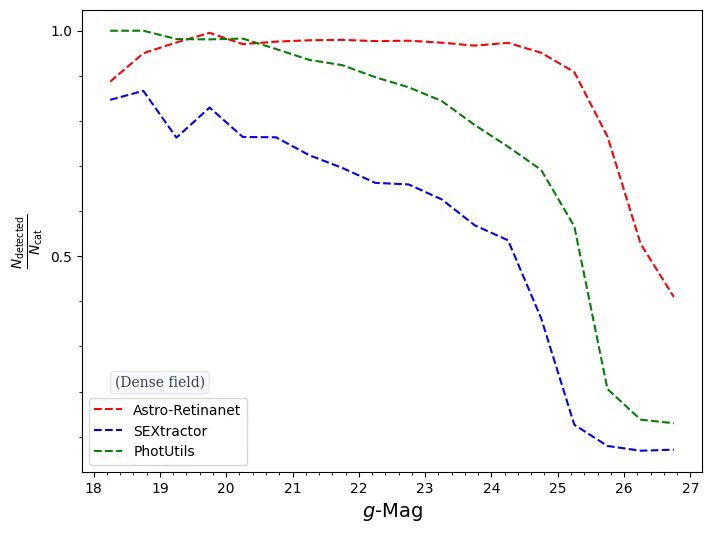}
\includegraphics[width=0.45\linewidth]{ 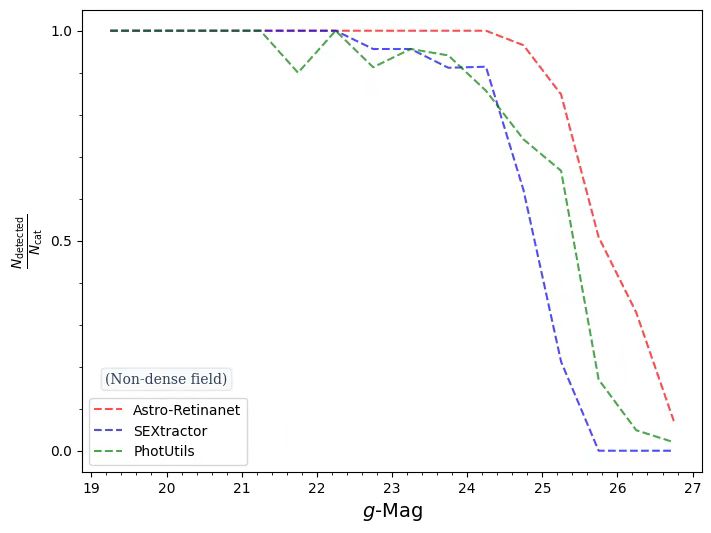}
\captionsetup{font=small} 
\caption{Detection completeness results of CSST simulated NGC 2298 with three method: Astro-RetinaNet in red dotted line, SExtractor in blue dotted line and Photutils in green dotted line. Left panel indicates the result of crowded stellar fields and right panel correspond to non-crowded stellar fields.}\label{fig:recover rate} 
\end{figure}

A substantial fraction of the errors arises from spurious detections in the flux wings of bright stars (mag $\leq 24$), where faint companions with large flux contrasts are difficult to distinguish from background fluctuations.
These false positives result in an average of 4–5 artifact detections per bright source and account for approximately $5.7\%$ of all detected signals in the simulated NGC~2298 field.

Considering the detection ability in the completeness result of Figure. \ref{fig:recover rate}.
In crowded stellar fields, our model demonstrates superior performance, maintaining $86.92\% $completeness for the bright end ($\rm mag < 25.5$), and approximately $43.8\%$ completeness at the faint end ($\rm mag >25.5$). 
In contrast, Photutils declines sharply from  $74.54\%$ at the bright end  to $29.37\%$ at the faint end, while SExtractor shows the poorest performance with an average completeness below 40\% across the full magnitude range.
Under sparse field conditions, all three methods perform comparably well for bright end, achieving nearly 99\% completeness.
However, their performance diverges significantly at the faint end: our model begins to decline at 25.5 mag, Photutils at 25 mag, and SExtractor as early as 24.5 mag. These results highlight our model's enhanced capability in both crowded and sparse field environments, particularly for faint source detection where traditional tools show substantial limitations.

\subsection{Result of M31}\label{subsec:M31 result}
% waitting for the new result\
This section evaluates the performance of the model on real M31 data \citep[PHAT,][]{M31_phat}.
We began by generating a simple PSF for the $F814W$ filter using TinyTim \citep{krist_tinytim}, a PSF modeling tool developed for the HST. Using the method described in section \ref{sec:method}, we then synthesized specific blended training data for $F814W$, incorporating a background noise model based on the CSST noise distribution. Finally, the trained model was applied to the M31 images. 
We adopt the PHAT catalog as a reference ground truth for performance evaluation, while noting that it represents the best-available catalog at the current observational resolution rather than an absolute ground truth in a strict sense.

\begin{figure}[htbp] 
\centering 
%\hspace{-0.4cm}
\includegraphics[width=0.95\linewidth]{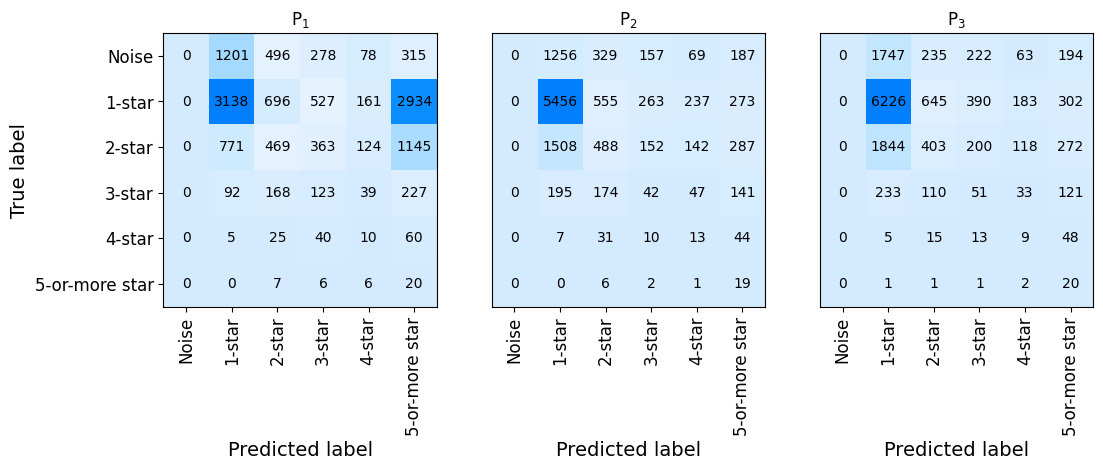}
\caption{Confusion matrices for the same M31 field obtained using the $\rm{P}_1,~\rm{P}_2$ and $\rm{P}_3$ models.
Increase or decrease the fraction of single star can lead more sensitive detection result, but more classification errors.}\label{fig:Matrix_m31_p123} 
\end{figure}

To evaluate the classification performance for blended sources, we applied the three single-star ratio distributions ($\rm{P}_1,~\rm{P}_2$ and $\rm{P}_3$), following the same strategy as for the CSST-simulated NGC 2298 data. 
The confusion matrices are shown in Figure. \ref{fig:Matrix_m31_p123}. The $\rm{P}_1$ distribution, which has the the lowest single-star precision of 42.1\% (3,138/7,456). In contrast, $\rm{P}_2$ and $\rm{P}_3$ achieve significantly higher and comparable single-star precision, 80.4\% (5,456/6,784) and 80.3\% (6,226/7,746), respectively.

For blended-source classification, $\rm{P}_3$ and $\rm{P}_2$ show similar performance. 
The numbers of spurious detections for $\rm{P}_1$, $\rm{P}_2$, and $\rm{P}_3$ are 2,368, 1,998, and 2,461, respectively, while the total numbers of detected objects are 11,156 ($\rm{P}_1$), 10,093 ($\rm{P}_2$), and 11,246 ($\rm{P}_3$).
For 2-star blends, the classification precision is 16.3\% (469/2,872), 18.9\% (488/2,577), and 14.2\% (403/2,837) for $\rm{P}_1$, $\rm{P}_2$, and $\rm{P}_3$, respectively. For 3-star blends, $\rm{P}_1$ reaches the highest precision at 18.9\% (123/649), whereas $\rm{P}_2$, and $\rm{P}_3$ are both below 10\%. 
Similar to the results obtained for the CSST-simulated NGC 2298 field, increasing the fraction of single stars in the training set leads to a higher rate of blended sources being misclassified as single stars. For the $\rm{P}_3$ model, 65.0\% (1844/2,837) of the 2-star blends and 35.9\% (233/648) of the 3-star blends are incorrectly classified as single stars.
Due to the limited number of higher-order blends, the corresponding results are not statistically significant.

\begin{figure}[htbp] 
\centering 
%\hspace{-0.4cm}
\includegraphics[width=0.7\linewidth]{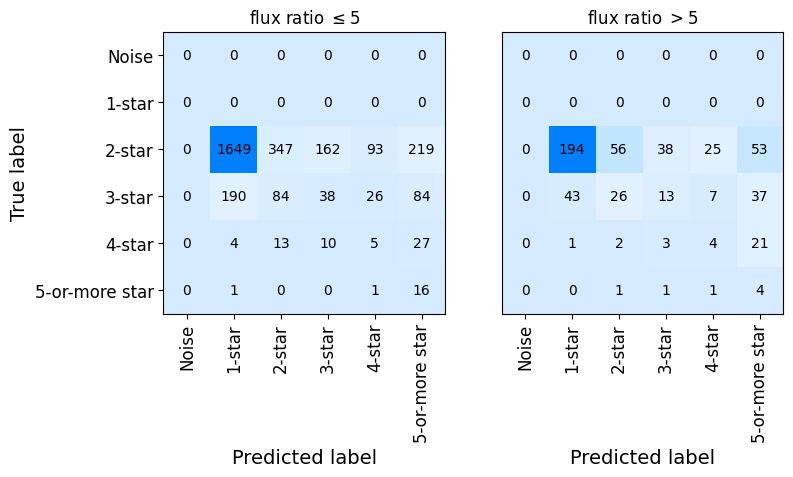}
\caption{ Confusion matrices for the classifications of M31 sources by the $\rm{P}_3$ model, separately for systems with flux contrasts greater than and less than a factor of 5. 
The subsample with flux contrasts below a factor of 5 dominates the population.
}\label{fig:HST_fluxCompare} 
\end{figure}

We further assess the impact of flux contrast on blending classification using the model trained with the $\rm{P}_3$ distribution. 
The blended samples are divided into two subsets according to whether the flux ratio between components is smaller or larger than 5 (Figure. \ref{fig:HST_fluxCompare}).

Low-contrast blends dominate the sample, comprising 84.8\% (2,969/3,499). For these systems, 66.8\% (1,649/2,470) of the 2-star blends are misclassified as single stars, compared to 53.0\% (194/366) in the high-contrast subset.
A similar behavior is found for the 3-star class: 45.0\% (190/422) of low-contrast blends and 34.1\% (43/126) of high-contrast blends are misclassified as single stars.

These results indicate that blends with small flux contrasts are more difficult to distinguish from single stars, whereas large-contrast blends are comparatively easier to identify.

In terms of stellar counts, Because the highly blended systems (4-star and 5-or-more-star) are rare and do not significantly affect the number statistics, we only quantify the impact of misclassification among the single-star, 2-star, and 3-star categories.
Misclassifying single stars as higher-order blends leads to an overestimation of 4,702 stars.
For the 2-star systems, 1,844 stars are underestimated, while 1,252 stars are overestimated.
For the 3-star systems, 576 stars are underestimated and 275 are overestimated.
Overall, misclassification into higher blended categories causes a net overestimation of 3,809 stars, corresponding to 24.5\% (3,809/15,549) of the detected sources.
However, when missed detections are taken into account, the model recovers only 70.2\% (17,817/25,396) of the stars listed in the PHAT catalog within the analyzed regions, which is about 14\% lower than the result for CSST-simulated NGC 2298.
This discrepancy is mainly due to the low fraction of highly blended systems (such as 4-star or 5-or-more-star) in M31 and the large fraction of faint sources ($\rm F814W > 27$) that are difficult for the model to detect.
Nevertheless, down to this limiting magnitude of $F814W \sim 27$, Astro-RetinaNet still detects 2,224 sources, outperforming Photutils and SExtractor by factors of 3.4 and 7.1, respectively.
As a result, although the model overestimates the star counts among the detected sources, the total number of stars in the full field is still systematically underestimated.
 
\begin{figure}[htbp] 
\centering 
%\hspace{-0.4cm}
\includegraphics[width=0.85\linewidth]{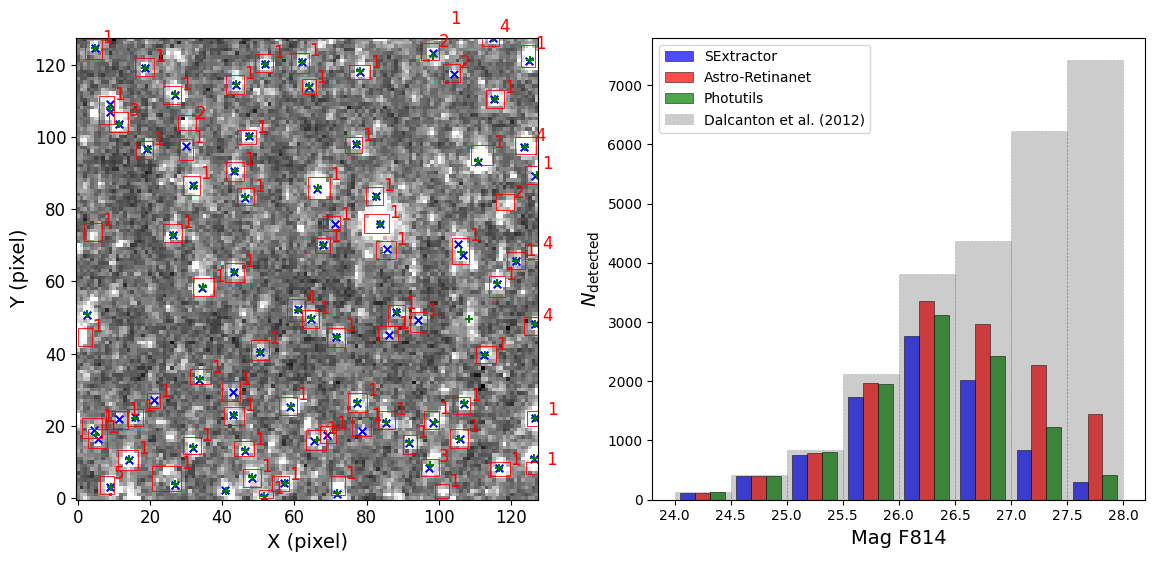}
\caption{Same as Figure. \ref{fig:Detection ability}, but for a subset of M31. Left: The detection result of Astro-RetinaNet, SExtractor and Photutils in a $500\times 500$ pixel region in entire test subset. Right: Magnitude distribution of test subset region after cross-matching with the PHAT catalog. Photutils performce better than our model in detections histogram. }\label{fig:M31_result} % \ref{fig:recover rate}.
\end{figure}

The classification and detection results of the blending are shown in Figure. \ref{fig:M31_result}. 
In the  context of the classification result, Astro-RetinaNet outperforms both SExtractor and Photutils, with its advantage growing for faint sources ($\rm{mag} > 26$).
As the magnitude increases, the completeness of Astro-RetinaNet decreases from 93.46\% in the $\rm{[26.0, 26.5]}$ bin to 26\% in the $\rm{[27.5, 28.0]}$ bin.
In comparison, Photutils and SExtractor drop from $84.24\%$ and $72.41\%$ to $7.8\%$ and $3.75\%$. 

However, the proposed model also exhibits several limitations compared with traditional source-extraction methods. 
For bright sources, SExtractor and Photutils can reliably locate the flux peak centers, whereas the model often produces multiple candidate detections around the peak, particularly in the PSF wings. 
For extremely bright sources approaching CCD saturation, the model may even fail to detect them, likely because such cases are not represented in the training set. 
Moreover, traditional tools are generally more robust in rejecting non-Gaussian background noise and rarely misclassify background fluctuations as real sources, while the model occasionally interprets such fluctuations as detections. 
This behavior suggests that the model relies primarily on local signal-to-noise thresholds and has not fully learned to distinguish true stellar PSF morphologies from noise patterns.

Several factors may contribute to the reduced performance of the model on the M31 test data compared to the simulation-based validation.

First, the background level in the M31 observations differs from that assumed in the simulations, which are based on the CSST instrumental configuration. This mismatch affects source detectability and classification, particularly for faint and blended objects.

Second, limitations in PSF modeling may also play a role. Although the $F814W$ PSF is generated using TinyTim, the PSF in real observations varies with the telescope’s optical state and exhibits spatial variations across the field, which are not fully captured by a single static PSF model.

Finally, differences in the magnitude distribution of single stars between simulations and observations have a significant impact. To avoid unrealistically faint blended components, we impose a lower magnitude limit of 24.5 mag for single star in the simulations. 
Consequently, in the M31 images, single stars fainter than this limit are more likely to be interpreted as blended systems, introducing a systematic bias in the classification of faint sources.

\section{Discussion}\label{discussion}
In the model performance analysis of section  \ref{Performance analysis}, 
the choice of a lower IoU threshold (0.2 compared to commonly used 0.5) accounts for our targets are predominantly small and frequently blended sources.
But this effect has little influence on star counts; only the source center live within the boundary, the source will be treated as ``detected''.

The current model operates on science-grade images corrected for instrumental effects, yet remains susceptible to residual noise—such as cosmic rays and diffraction spikes—which shad negative affects on detection and classification. 
Moreover, the model defines the detection depth based on post-detection confidence scores rather than traditional SNR. This approach explains both the misclassification of noise as real signals and the omission of faint sources. 
Addressing these issues requires either enhancing the model's ability to learn noise characteristics or improving preprocessing techniques to more effectively clean such artifacts.

It is also critical to acknowledge that the current evaluation reflects the model’s proficiency in handling the simulation conditions, and its performance on real CSST data may differ due to the gaps between the simulations and actual observational conditions (e.g., unmodeled systematics, cosmic ray impacts or variations in PSF). 
In future work, a dedicated investigation of domain adaptation strategies and real-data validation will be addressed, as these topics extend beyond the scope of this methodological study.
As for other telescope survey data, because of various PSF profile, the transfer learning method is needed to improve the model's learning efficiency.

\section{Conclusions}\label{sec:conclusion}
In this work, we develop a new Machine-learning method, Astro-RetinaNet, for detecting and classifying blending stars in astronomical images. 
Astro-RetinaNet is a fast and accurate algorithm that mainly handle star counts and blending stars classification in crowded stellar fields,  the detection performance of our model is statistically consistent with that of SExtractor and Photutils.
In a  highly blending field, it can identify the blending sources and estimate a rough number of components. 
Results from simulated NGC 2298 data and real HST observed M31 show that Astro-RetinaNet recovers approximately 84\% of the stellar counts in NGC 2298, whereas the recovery rate decreases to about 70\% for M31, primarily due to the limited number of highly blended sources in the sample.
With fine-tuning of the PSF profile according to the observed crowded stellar fields, we can then give a $\pm 1$ uncertainty in predicting the blending number.

Our validation tests show that the model achieves precision  ($\bm{\frac{TP}{TP+FP}}$) of $96.3\%$ (18,601/19,325, with $TN =31,801$) for the single-star class, with performance gradually declining as the complexity of blending increases.
The classification precision rates are: $89.3\%$ (8,210/9,196, $TN = 40,998$)  for 2-star, $69.9\%$ (5,672/8,112, $TN = 41,293$) for 3-star, $50.4\%$ (3,838/7,617, $TN = 41,440$) for 4-star, and $75.2\%$ (5,348/7,113, $TN = 42,821$) for 5-or-more-blending systems.
We think that the wild increase in 5-or-more-blending units comes from its blending definition, which consists of all higher-order blending units ($\geq5~\rm stars$). 

We further test our model using CSST-simulated NGC 2298 data and real observations of M31 to analyze its performance in identifying blended sources. 
We then examined how different single-star proportion distributions ($\rm{P}_1 = (1:2:2:2:2), \quad \rm{P}_2 = (1:1:1:1:1), \quad \rm{P}_3 = (2:1:1:1:1).$) in the training set affect the classification of blended sources.
We find that adjusting the single-star fraction—either decreasing or increasing it—primarily changes the model’s sensitivity to detected signals, but does not change its tendency to misclassify blended systems as single stars.
While the higher single-star fraction in $\rm{P}_3$ improves the precision of the single-star class (from 75.9\% (7,377/9,716, $TN = 3,710$) to 82.3\% (8,580/10,419, $TN = 2,346$) for CSST-simulated NGC 2298, and from 42.1\% (3,138/7,456, $TN =2,832$) to 80.3\% (6,226/7,746, $TN = 1,417$) for HST-observed M31 field), it also increases the model’s tendency to misclassify blended sources as single stars. 
In the CSST-simulated NGC 2298 field,  71.0\% (2,603/3,659) of 2-star systems, 57.1\% (776/1358) of 3-star systems, 42.1\% (226/537) of 4-star systems, and 24.2\% (127/524) of systems with 5-or-more-star are still incorrectly labeled as single stars.
Similarly, in the HST-observed M31 field, 65.0\% (1,844/2,837) of the 2-star blends and 35.9\% (233/648) of the 3-star blends are misclassified as single stars.
Due to the limited number of higher-order blends in M31, the corresponding results are not statistically significant.

We also compare its detection ability with the traditional algorithm SExtracotr and Photutils. Our model demonstrated better performance both in the crowded and non-crowded field. 
In the  crowded field of NGC 2298, Astro-RetinaNet can recall the $86.92\%$  bright end ($\rm mag < 25$) and the $43.8\%$ completeness for the faint end ($\rm mag \geq 25$).
In M31, Astro-RetinaNet detects 2,224 sources down to a limiting magnitude of $F814W \sim 27$ and outperforms Photutils and SExtractor by factors of 3.4 and 7.1.
Despite its overall performance, the model remains less robust than traditional methods for very bright sources and in the presence of non-Gaussian background noise.
In particular, it tends to produce spurious detections in PSF wings, fails near saturation, and occasionally misclassifies background fluctuations as real sources, indicating an incomplete learning of true PSF morphologies.

In the future work, we will use a Bayesian model to determine the position and magnitude of each component of the mixing source, then it can also be used to analyze stellar population of the host galaxy, which can be of great help in studying the evolutionary history of galaxies and in searching for faint dwarf galaxies. The related papers are in preparation.

\section{Acknowledgements}
The authors acknowledge the support by the China Manned Space Project (No. CMS-CSST-2025-A21), the National Key Research and Development Program of China (2025YFF0511000) and the National Natural Science Foundation of China (12433012, 12373097). This work is also supported by the National Astronomical Observatories of the Chinese Academy of Sciences, under Project No. E4TQ680101. 

The CSST related mock data in this work are created by the CSST Simulation Team, which is supported by the CSST scientific data processing and analysis system of the China Manned Space Project. 
\bibliography{bibref}{}
\bibliographystyle{aasjournal}
\end{document}